\documentclass[11pt]{iopart}

\usepackage{iopams}
\usepackage{amsfonts}
\usepackage{amsbsy} 
\usepackage{epsfig}
\usepackage{bbm}
\usepackage{latexsym}

\usepackage{color}
\usepackage[breaklinks]{hyperref}

\def\ben{\begin{equation}}
\def\een{\end{equation}}
\def\bena{\begin{eqnarray}}

\def\eena{\end{eqnarray}}

\def\bR{\mathbb R}
\def\bC{\mathbb C}

\def\b1{e^0}

\def\im{{\rm i}}

\def\half{\frac{1}{2}}
\def\SU{{\rm SU}}
\def\SO{{\rm SO}}

\def\com{\color{magenta}}

\newcommand{\oarX}[1]{\href{http://arxiv.org/abs/#1}{{\ttfamily\com #1}}}
\newcommand{\arX}[1]{\href{http://arxiv.org/abs/#1}{{\ttfamily\com arXiv:#1}}}

\begin{document}

\begin{flushright}\end{flushright}

\title{Quantum cosmology of (loop) quantum gravity condensates: An example}

\author{Steffen Gielen}
\address{Perimeter Institute for Theoretical Physics, 31 Caroline St. N., Waterloo, Ontario N2L 2Y5, Canada}
\ead{sgielen@perimeterinstitute.ca}

\begin{abstract}
Spatially homogeneous universes can be described in (loop) quantum gravity as condensates of elementary excitations of space. Their treatment is easiest in the second-quantised group field theory formalism which allows the adaptation of techniques from the description of Bose--Einstein condensates in condensed matter physics. Dynamical equations for the states can be derived directly from the underlying quantum gravity dynamics. The analogue of the Gross--Pitaevskii equation defines an anisotropic quantum cosmology model, in which the condensate wavefunction becomes a quantum cosmology wavefunction on minisuperspace. To illustrate this general formalism, we give a mapping of the gauge-invariant geometric data for a tetrahedron to a minisuperspace of homogeneous anisotropic 3-metrics. We then study an example for which we give the resulting quantum cosmology model in the general anisotropic case and derive the general analytical solution for isotropic universes. We discuss the interpretation of these solutions. We suggest that the WKB approximation used in previous studies, corresponding to semiclassical fundamental degrees of freedom of quantum geometry, should be replaced by a notion of semiclassicality that refers to large-scale observables instead. 
\end{abstract}


\maketitle

\section{Introduction}

There is by now a variety of approaches to the problem of quantum gravity which are actively pursued \cite{danielebook}. Research into any of these directions generally addresses one of two basic aims. The first is to show that a proposed theory of quantum gravity is in itself consistent and that its objects are mathematically well-defined, computable, and can be translated into observable quantities, so that the theory can, at least in principle, be confronted with experiment. The second aim is a derivation of the phenomenology of the theory, which usually requires taking a `low-energy' or `semiclassical' regime, in which the theory should at least be consistent  with present observational constraints on deviations from the predictions of general relativity and the standard model of particle physics. It is then often claimed that any genuine quantum-gravitational effect, going beyond separate predictions of general relativity or the standard model, would be intrinsically unobservable, since the Planck scale is many orders of magnitude above the energy scales probed in particle accelerators or hypothetical experiments. However, while it is indeed difficult to come up with present-day experiments that probe Planck-scale physics (for some efforts in this direction, see \cite{giovanni}), the very early universe provides a natural laboratory in which quantum gravity effects can be expected to play a role. 

Inflation, the standard paradigm for the physics of the very early universe, has been spectacularly corroborated in the recent observations made by Planck \cite{planck} and BICEP2 \cite{bicep}. However, despite its phenomenological success, there are several theoretical issues that remain open: the inflaton and its potential are not part of the standard model, and have to be added by hand. While inflation provides a picture in which the physics at the Big Bang singularity is not observationally relevant today, as its imprint has been stretched outside the causal horizon during the accelerated expansion, theorems such as \cite{theorem} show that inflationary spacetimes have a past singularity, so that there is still a need for a more complete theory. Eternal inflation seems to have drastic and contentious theoretical consequences \cite{eternal}. Observationally, the BICEP2 results seem to imply a violation of the Lyth bound \cite{lyth}: the inflaton field presumably varies over super-Planckian scales during inflation. All of this motivates the study of quantum-gravitational models with regard to their predictions for cosmology.

The spacetimes relevant for cosmology are to a very good approximation spatially homogeneous. One can use this fact and perform a symmetry reduction of the classical theory (general relativity coupled to a scalar field or other matter) assuming spatial homogeneity, followed by a `quantisation' of the reduced system. Inhomogeneities are usually added perturbatively. This leads to models of quantum cosmology \cite{admrefs} which can be studied on their own, without the need for a manageable full theory of quantum gravity. While this approach can be pursued with profit to some extent, and is claimed to make potentially observable predictions \cite{kiefercmb}, there is no unambiguous interpretation of calculations that supposedly result from truncation of an unknown underlying theory. For instance, since one is generally ignorant about the physical inner product in full quantum gravity, the predictive power of computing wavefunctions is not clear.

Loop quantum gravity (LQG) has some of the structures one would expect in a full theory of quantum gravity: kinematical states corresponding to functionals of the Ashtekar--Barbero connection can be rigorously defined, and geometric observables such as areas and volumes are well-defined as operators, typically with discrete spectrum \cite{LQGbook}. Using the LQG formalism in quantising symmetry-reduced gravity leads to loop quantum cosmology (LQC) \cite{LivRev}. Because of the structures of LQG, LQC allows a rigorous analysis of issues that could not be addressed within the Wheeler--DeWitt quantisation of conventional quantum cosmology, such as a definition of the physical inner product. Recently, LQC has made contact with CMB (cosmic microwave back-ground) observations, as the usual inflationary scenario is now discussed in LQC \cite{abhayagullo}.

One missing ingredient in the formalism of LQC is its embedding into the full setting of LQG. Just as in conventional quantum cosmology, one has performed a symmetry reduction before quantisation, and truncated almost all degrees of freedom present in the full Hilbert space of LQG. A different approach aiming at a more complete picture would be to work within the full Hilbert space, identify states that can represent macroscopic, (approximately) spatially homogeneous universes, and extract information about their dynamics. Clearly, this last step will involve many approximations, but since these are approximations for equations of the full theory, one has some control about the error made. Already the identification of suitable states that represent cosmological spacetimes is challenging in a theory like LQG: because of the notion of background independence built into the definition of the theory, the most natural notion of vacuum state is the `no space' state, which has zero expectation value for geometric observables (areas, volumes, etc). Elementary excitations over this vacuum are usually interpreted as distributional geometries, and a macroscopic nondegenerate configuration is unlikely to be found as a small perturbation of this vacuum. 

A new approach towards addressing the issue of how to describe cosmologically relevant universes in (loop) quantum gravity was recently proposed in \cite{PRL,JHEP} \footnote{For some alternative approaches towards the same problem, see \cite{uther}.}. This proposal uses the group field theory (GFT) formalism, itself a {\em second quantisation} formulation of the kinematics and dynamics of LQG \cite{2ndq}: one has a Fock space of LQG spin network vertices (or tetrahedra, as building blocks of a simplicial complex), annihilated and created by the field operator $\hat\varphi$ and its Hermitian conjugate $\hat\varphi^{\dagger}$, respectively. The advantage of using this reformulation is that field-theoretic techniques are available, as a GFT is a standard quantum field theory on a curved (group) manifold ({\em not} to be interpreted as spacetime). In particular, one can define {\em coherent} or {\em squeezed} states for the GFT field, analogous to states used in the physics of Bose--Einstein condensates or in quantum optics; these represent {\em quantum gravity condensates}. They describe a large number of degrees of freedom of quantum geometry in the same microscopic quantum state, which is the analogue of homogeneity for a differentiable metric geometry. This idea was made explicit in \cite{JHEP}: after embedding a condensate of tetrahedra into a smooth manifold representing a spatial hypersurface, one shows that the spatial metric (in a fixed frame) reconstructed from the quantum state is compatible with spatial homogeneity. As the number of tetrahedra is taken to infinity, a continuum homogeneous metric can be approximated to a better and better degree.

At this stage, the condensate states defined in this way are kinematical. They are gauge-invariant (locally Lorentz invariant) by construction, and represent geometric data invariant under (active) spatial diffeomorphisms, but they do not satisfy any dynamical equations corresponding to a Hamiltonian constraint in geometrodynamics. The strategy followed in \cite{PRL, JHEP} for extracting information about the dynamics of these states is the use of Schwinger--Dyson equations of a given GFT model. These give constraints on the $n$-point functions of the theory evaluated in a given condensate state (approximating a non-perturbative vacuum), which can be translated into differential equations for the `condensate wavefunction' used in the definition of the state. Again, this is analogous to condensate states in many-body quantum physics, where such an expectation value gives, in the simplest case, the Gross--Pitaevskii equation for the condensate wavefunction. The truncation of the infinite tower of such equations to the simplest ones is part of the approximations made. As argued in \cite{PRL, JHEP}, the effective dynamical equations thus obtained can be viewed as defining a {\em quantum cosmology model}, with the condensate wavefunction interpreted as a quantum cosmology wavefunction. This provides a general procedure for deriving an effective cosmological dynamics directly from the underlying theory of quantum gravity. In a specific example, it was shown how a particular quantum cosmology equation of this type, in a semiclassical WKB limit and for isotropic universes, reduces to the classical Friedmann equation of homogeneous, isotropic universes in general relativity.

The purpose of this paper, apart from reviewing the formalism introduced in detail in \cite{JHEP}, is to analyse more carefully the quantum cosmological models derived from quantum gravity condensate states in GFT. In particular, the formalism identifies the gauge-invariant configuration space of a tetrahedron with the minisuperspace of homogeneous (generally anisotropic) geometries. We will justify this interpretation and propose a convenient set of variables for the gauge-invariant geometric data, which can be mapped to the variables of a general anisotropic Bianchi model (for which the metric is not diagonalised and has six components). We will then revisit the example that led to the Friedmann equation in \cite{PRL, JHEP} and study it directly as a quantum cosmology equation, without a WKB limit. The Friedmann equation arising in a WKB limit in \cite{PRL} appeared to have no solutions, as there was a mismatch between the curvature of the gravitational connection, assumed to be small on the scale of the tetrahedra, and the spatial curvature term which was large on the same scale. Here we find simple solutions to the full quantum equation, corresponding to isotropic universes. They can only satisfy the condition of rapid oscillation of the WKB approximation for large positive values of the coupling $\mu$ in the GFT model. For $\mu<0$, states are sharply peaked on small values for the curvature, describing a condensate of near-flat building blocks, but these do not oscillate. This supports the view that rather than requiring semiclassical behaviour at the Planck scale, semiclassicality should be imposed only on large-scale observables.

\section{From quantum gravity condensates to quantum cosmology}

Here we review the relevant steps in the construction of effective quantum cosmology equations for quantum gravity condensates. We work in the group field theory (GFT) formalism, which is a second quantisation formulation of loop quantum gravity spin networks (of fixed valency), or their dual interpretation as simplicial geometries. For full details of the precise relation between the two, see \cite{2ndq}.

The basic structures of the GFT formalism in four dimensions are a complex-valued field $\varphi:G^4\rightarrow\bC$, satisfying a gauge invariance property
\ben
\varphi(g_1,\ldots,g_4)=\varphi(g_1h,\ldots,g_4h)\quad \forall h\in G\,,
\label{gauge}
\een
and the basic (non-relativistic) commutation relations imposed in the quantum theory
\ben
[\hat\varphi(g_I),\hat\varphi^{\dagger}(g_I')]={\bf 1}_G(g_I,g'_I)\,,\quad [\hat\varphi(g_I),\hat\varphi(g_I')]=[\hat\varphi^{\dagger}(g_I),\hat\varphi^{\dagger}(g_I')]=0\,.
\label{comm}
\een
The relations (\ref{comm}) are analogous to those of non-relativistic scalar field theory, where the mode expansion of the field operator defines annihilation operators, $\hat\phi(\vec{x})=\sum_{\bf k}\hat{a}_{{\bf k}}\phi_{{\bf k}}(\vec{x})$, and similarly for the Hermitian conjugate $\hat\phi^{\dagger}(\vec{x})=\sum_{\bf k}\hat{a}^{\dagger}_{{\bf k}}\overline{\phi_{{\bf k}}}(\vec{x})$. In GFT, the domain of the field(s) is four copies of a Lie group $G$, interpreted as the local gauge group of gravity, which can be taken to be $G={\rm Spin}(4)$ for Riemannian and $G={\rm SL}(2,\bC)$ for Lorentzian models. In loop quantum gravity, the gauge group is the one given by the classical Ashtekar--Barbero formulation, $G=\SU(2)$. The property (\ref{gauge}) encodes invariance under gauge transformations acting on spin network vertices, as we will see shortly. In (\ref{comm}), ${\bf 1}_G$ is an identity operator on the group compatible with (\ref{gauge}). For compact $G$, 
\ben
{\bf 1}_G(g_I,g'_I) = \int_G {\rm d}h \;\prod_{I=1}^4 \delta(g_I h (g'_I)^{-1})\,,
\een
where here and in the following the measure ${\rm d}h$ is normalised to $\int{\rm d}h=1$.

One then defines a Fock vacuum $|\emptyset\rangle$ annihilated by all $\hat\varphi(g_I)$, analogous to the diffeomorphism-invariant Ashtekar--Lewandowski vacuum of LQG, with zero expectation value for all area or volume operators. The conjugate $\hat\varphi^{\dagger}(g_I)$ acting on $|\emptyset\rangle$ creates a GFT `particle', interpreted as a 4-valent spin network vertex or a dual tetrahedron:
\ben
\begin{picture}(320,100)
\put(20,50){$\hat\varphi^{\dagger}(g_1,g_2,g_3,g_4)|\emptyset\rangle=|$}\put(270,50){$\rangle$}
\put(202,52){$\bullet$}\put(200,100){\line(-1,-3){30}}\put(170,10){\line(3,1){90}}\put(260,40){\line(-1,1){60}}\put(170,10){\line(-1,1){30}}\put(140,40){\line(1,1){60}}
\put(205,55){\line(-1,0){70}}\put(205,55){\line(1,1){50}}\put(205,55){\line(1,-1){50}}\put(205,55){\line(-1,-3){20}}
\put(170,60){$g_1$}\put(225,90){$g_2$}\put(225,40){$g_3$}\put(185,30){$g_4$}
\end{picture}
\label{particle}
\een
The geometric data attached to this tetrahedron, four group elements $g_I \in G$, is interpreted as parallel transports of a (gravitational) connection along links dual to the four faces. Gauge transformations act on the vertex where these links meet as $g_I\mapsto g_I h$, which is the reason for requiring (\ref{gauge}).

The LQG interpretation of (\ref{particle}) is that of a state that fixes the parallel transports of the Ashtekar--Barbero connection to be $g_I$ along the four links given by the spin network, while they are undetermined everywhere else. Again, this is analogous to the Fock space of usual scalar field theory in which $|\vec{x}\rangle=\hat\phi^{\dagger}(\vec{x})|0\rangle$ defines a particle at position $\vec{x}$.

In the canonical formalism of Ashtekar and Barbero, the canonically conjugate variable to the connection is a {\em densitised (inverse) triad}, with dimensions of area, that encodes the spatial metric. The GFT formalism can be translated into this `momentum space' formulation by use of a {\em non-commutative Fourier transform} \cite{noncf},
\ben
\tilde\varphi(B_1,\ldots,B_4)=\int ({\rm d} g)^4\;\prod_{I=1}^4 e_{g_I}(B_I)\; \varphi(g_1,\ldots,g_4)
\een
where $e_{g_I}(B_I)$ is a choice of plane wave on $G$. Since $G$ is non-Abelian, the product of plane waves defined by $e_g(B)\star e_{g'}(B)=e_{gg'}(B)$ is non-commutative; its extension to general superpositions of plane waves turns the space parametrised by $B_I$ into a non-commutative geometry, which is the Lie algebra $\mathfrak{g}^{\oplus 4}$ of $G^4$. 

The geometric interpretation of the variables $B_I\in\mathfrak{g}$ is as geometric bivectors associated to a spatial triad $e$, defined by the integral $\int_{\triangle_I} e^A\wedge e^B$ over a face $\triangle_I$ of the tetrahedron. Hence, the one-particle state
\ben
|B_1,\ldots,B_4\rangle = \hat{\tilde\varphi}(B_1,\ldots,B_4)|\emptyset\rangle
\een
defines a tetrahedron with minimal uncertainty in the `fluxes', {\em i.e.} oriented area elements $\int_{\triangle_I} e^A\wedge e^B$, given by $B_I$. \footnote{The variables $g_I$ and $B_I$ should be thought of as invariant under active (spatial) diffeomorphisms, by construction. However, in canonical gravity
there are gauge transformations which represent the passive version of diffeomorphisms. One should be able to identify an action of such transformations on the GFT variables. In three-dimensional GFT, this has been done in \cite{gftdiff}.} Again, in the LQG interpretation this state completely determines the metric variables for one tetrahedron, while being independent of all other degrees of freedom of geometry in a spatial hypersurface.

The idea of {\em quantum gravity condensates} is to use many excitations over the Fock space vacuum $|\emptyset\rangle$, all in the same microscopic configuration, to better and better approximate a smooth homogeneous metric (or connection), as a many-particle state can contain information about the connection and the metric at many different points in space. Choosing this information such that it is compatible with a spatially homogeneous metric while leaving the particle number $N$ free, the limit $N\rightarrow\infty$ corresponds to a continuum limit in which a homogeneous metric geometry is recovered.

In the simplest case, the definition for GFT condensate states is
\ben
|\sigma\rangle := \mathcal{N}(\sigma) \exp\left(\hat\sigma\right)|\emptyset\rangle \quad{\rm with}\quad \hat\sigma := \int ({\rm d} g)^4\; \sigma(g_I)\hat\varphi^{\dagger}(g_I) \,,
\label{singcon}
\een
where $\mathcal{N}(\sigma)$ is a normalisation factor. The exponential creates a coherent configuration of many building blocks of geometry. At fixed particle number $N$, a state of the form $\hat\sigma^N|\emptyset\rangle$ would be interpreted as defining a metric (or connection) that looks spatially homogeneous when measured at the $N$ positions of the tetrahedra, given an embedding into space. However, one does not work at fixed particle number, but there is a sum over all possible particle numbers. The condensate picture is rather different from many constructions in the literature: it does not use a fixed graph or discretisation of space.

The above summary gives an intuitive picture rather than full details, which can be found in \cite{JHEP}. It uses the geometric interpretation of LQG spin network states, which is obtained by viewing LQG as a quantisation of a classical action for general relativity. Ultimately, the identification of the degrees of freedom of the quantum theory with classical geometric quantities involves a detailed understanding of the continuum limit, which is largely an open issue \cite{ditt}. Computing an effective dynamics for the reconstructed macroscopic `metric', and verifying whether it satisfies Einstein's equations (with higher curvature corrections), would be an important step in this direction. GFT condensates can address this question in the case of spatial homogeneity.

While spatial homogeneity requires that all elementary building blocks of geometry are in the same microscopic configuration, it does not state what the elementary building blocks are. A natural second type of condensate is a condensate of `molecules' of two tetrahedra, with pairwise identified faces. It is defined by
\ben
|\xi\rangle := \mathcal{N}(\xi) \exp\left(\hat\xi\right)|0\rangle\,, \quad \hat\xi := 
\half\int ({\rm d} g)^4 ({\rm d} h)^4\; \xi(g_I^{-1} h_I)\hat\varphi^{\dagger}(g_I)\hat\varphi^{\dagger}(h_I)\,.
\label{dipolestate}
\een
In terms of LQG spin networks, the elementary building block of (\ref{dipolestate}) is a `dipole' graph for which the four links going out of one vertex all meet at a second vertex, thus forming a gauge-invariant closed spin network. Indeed, using (\ref{gauge}), the condensate wavefunction $\xi$ in (\ref{dipolestate}) is separately invariant under {\em two} gauge transformations,
\ben
\xi(g_1,\ldots,g_4)=\xi(k g_1 k',\ldots,k g_4 k')\quad \forall\, k,k'\in G\,.
\een
These transformations are local gauge transformations in the geometric interpretation of the GFT variables, acting respectively on the vertex of the tetrahedron in (\ref{particle}) and on its boundary (contracted to a second vertex for the dipole). In terms of the dual Lie algebra variables, the first type of transformation means that the bivectors add to zero, $\sum_{I=1}^4 B_I=0$, while the second one is a gauge transformation $B_I\mapsto kB_Ik^{-1}$.

In order to only depend on geometric variables and not on a local choice of Lorentz frame, the condensate must be invariant under both sets of transformations. Hence, in the case of (\ref{singcon}), we impose that $\sigma(g_I)=\sigma(k\,g_I)\;\forall k\in G$.

In both cases, the GFT condensate is defined in terms of a wavefunction on $G^4$ invariant under separate left and right actions of $G$ on $G^4$. The strategy introduced in \cite{JHEP} is then to demand that the condensate solves the GFT quantum dynamics, expressed in terms of the Schwinger--Dyson equations which relate different $n$-point functions for the condensate. An important approximation is to only consider the simplest Schwinger--Dyson equations, which will give equations of the form
\ben
\left(\hat{\mathcal{K}}\sigma\right)(g_1,\ldots,g_4)+\left(\hat{\mathcal{V}}\sigma\right)(g_1,\ldots,g_4)=0\,,
\label{effeq}
\een
where $\hat{\mathcal{K}}$ is a linear (potentially nonlocal) differential operator, and $\hat\mathcal{V}\sigma$ can be a nonlinear, nonlocal functional of $\sigma$ and $\bar\sigma$, the two terms coming from the kinetic and potential terms in the GFT action. This is again analogous to the case of the Bose--Einstein condensate where the simplest equation of this type (the expectation value of the classical equation of motion) gives the Gross--Pitaevskii equation.

In the case of a real condensate, the condensate wavefunction $\Psi(\vec{x})$, corresponding to a nonzero expectation value of the field operator, has a direct physical interpretation: expressing it in terms of amplitude and phase, $\Psi(\vec{x})=\sqrt{\rho(x)}\,e^{-\im \theta(\vec{x})}$, one can rewrite the Gross--Pitaevskii equation to discover that $\rho(x)$ and $\vec{v}(x)=\nabla \theta(\vec{x})$ satisfy {\em hydrodynamic} equations in which they correspond to the density and the velocity of the quantum fluid defined by the condensate. Microscopic quantum variables and macroscopic classical variables are directly related.

The wavefunction $\sigma(g_I)$ or $\xi(g_I)$ of the GFT condensate should play a similar role. It is not just a function of the geometric data for a single tetrahedron, but equivalently a function on a {\em minisuperspace} of spatially homogeneous universes. The effective dynamics for it, extracted from the fundamental quantum gravity dynamics given by a GFT model, can then be interpreted as a quantum cosmology model. The resulting quantum cosmology equations are in general nonlinear, which extends the usual formalism of Schr\"odinger-type linear equations but has been proposed in a different context before \cite{nonlin}. In the rest of this paper, we will make the interpretation of these equations as quantum cosmology models more explicit, and study a concrete example.

\section{Minisuperspace = gauge-invariant configuration space of a tetrahedron}

Condensate states of the type discussed are determined by a wavefunction $\sigma$, which is a complex-valued function on the space of four group elements (for given gauge group $G$) which is invariant under
\ben
\sigma(g_1,\ldots,g_4)=\sigma(kg_1k',\ldots,kg_4k')\,,\quad k,k'\in G\,,
\label{sigmainv}
\een
and hence really a function on $G\backslash G^4/G$. This quotient space is a smooth manifold with boundary, without a group structure. It is the gauge-invariant configuration space of the geometric data associated to a tetrahedron or, perhaps more naturally, of a `dipole' configuration of two tetrahedra with pairwise identified faces. When the effective quantum dynamics of GFT condensate states is reinterpreted as (perhaps nonlinear) quantum cosmology equations, $G\backslash G^4/G$ becomes a {\em minisuperspace} of spatially homogeneous geometries. 

For consistency, the dynamics given by $\hat\mathcal{K}$ and $\hat\mathcal{V}$ in (\ref{effeq}) must be compatible with the symmetries of $\sigma$, given by the left and right action of $G$ on $G^4$,
\ben
[\hat\mathcal{K},L_k]=[\hat\mathcal{K},R_{k'}]=0\,,\quad k,k'\in G\,,
\label{kinetic}
\een
and similar for $\hat\mathcal{V}$. These operators then act on the Hilbert space of condensate wavefunctions defined on $G\backslash G^4 / G$.

To proceed, we note that there is a natural bijection of quotient spaces,
\ben
\beta:\;G\backslash G^4/G \rightarrow G^3/{\rm Ad}_G\,,\quad [g_1,g_2,g_3,g_4]\mapsto [g_1g_4^{-1},g_2g_4^{-1},g_3g_4^{-1}]\,,
\een
with inverse
\ben
\beta^{-1}:\; G^3/{\rm Ad}_G\rightarrow G\backslash G^4/G\,,\quad [g_1,g_2,g_3]\mapsto [g_1,g_2,g_3,e]\,,
\een
where ${\rm Ad}_G$ is the adjoint action of $G$ on $G^3$ which maps $g_i\mapsto kg_ik^{-1}$. Hence one can equivalently view $\sigma$ as a function on $G^3/{\rm Ad}_G$. Its non-commutative Fourier transform
\ben
\tilde\sigma(B_1,B_2,B_3)=\int({\rm d} g)^3 \;\sigma(g_1,g_2,g_3)\prod_{i=1}^3 e_{g_i}(B_i)
\label{fourier}
\een
satisfies $\tilde\sigma(B_1,B_2,B_3)=\tilde\sigma(kB_1k^{-1},kB_2k^{-1},kB_3k^{-1})$ for all $k\in G$, due to the property $e_{kgk^{-1}}(B)=e_g(k^{-1}Bk)$ of the plane waves \cite{noncf}, and is thus a function on $\mathfrak{g}^{\oplus 3}/{\rm Ad}_G\equiv ({\rm Lie}(G^3))/{\rm Ad}_G$. This latter quotient space is closely related to the space of homogeneous spatial metrics: any homogeneous metric is specified by giving a group action on a manifold, and fixing the metric at one point in the manifold. Focussing on non-degenerate metrics, spatially homogeneous metrics are in one-to-one correspondence to elements of the homogeneous space ${\rm GL(3)}/ {\rm O}(3)\simeq {\rm SL(3)}/ {\rm O}(3)\times(\bR\backslash\{0\})$ (see {\em e.g.} \cite{super} for this and more general properties of the superspace of 3-metrics).

As a vector space, $\mathfrak{g}^{\oplus 3}$ is just $\bR^{{\rm dim}(G)\times 3}$. Choosing $G=\SU(2)$ and assuming the non-degeneracy condition ${\rm Tr}(B_1 B_2 B_3)\neq 0$ means restricting to the subspace ${\rm GL}(3)/{\rm Ad}_{\SU(2)}\subset \bR^{3\times 3}/{\rm Ad}_{\SU(2)}$. The orbits of the action $\SU(2)$ on this space are smaller than the orbits of ${\rm O}(3)$, as they preserve the sign of the invariant ${\rm Tr}(B_1 B_2 B_3)$. Restricting to ${\rm Tr}(B_1 B_2 B_3)>0$, {\em i.e.} making a choice of orientation, the domain of $\sigma$ in (\ref{fourier}) is indeed just the space of (non-degenerate) homogeneous 3-metrics.

The above is not just a topological identification of quotient spaces, but follows from the geometric interpretation of the GFT data attached to tetrahedra, or pairs of tetrahedra. As anticipated above, the Lie algebra elements $B_I$ are interpreted as the discretised analogue of a triad of 1-forms $e^A$,
\ben
B^{AB}_I \sim \int_{\triangle_I} e^A \wedge e^B\,.
\een
One of the assumptions of GFT condensates is that the discrete $B^{AB}_I$ are a good approximation to a continuum homogeneous metric, which can then be reconstructed from the geometric data in the GFT states. One hence assumes the reconstructed geometry to be almost constant over the scale of the tetrahedra, so that one can define
\ben
B^{AB}_i =: {\epsilon_i}^{jk}e^A_j e^B_k
\label{def}
\een
with the $e^A_j$ defining a `triad' at a given reference point ({\em e.g.} one of the vertices) of a tetrahedron. Assuming nondegeneracy, the space of such `triads' is ${\rm GL}(3)$, and its gauge-invariant data corresponds to elements of ${\rm GL(3)}/ {\rm O}(3)$; the $B_i$ then simply correspond to the densitised inverse triad of LQG.

If the GFT gauge group $G$ is chosen to be a four-dimensional rotation group such as ${\rm SL}(2,\bC)$, the identification of variables is more subtle. (\ref{def}) is no longer a definition of $e^A_i$, as there need not be a set of vectors $e^A_i$ for given $B_i\in\mathfrak{g}$ so that (\ref{def}) holds. The restriction to {\em simple} $B_i$ of the form (\ref{def}) must be ensured by imposing {\em simplicity constraints} which restrict the modes appearing in the Peter--Weyl expansion of the GFT field in representations of $G$ (for a review of various prescriptions in the spin foam language, see \cite{sfsimp}). In the GFT formalism, this can be done at the level of the field itself, or in the action, either in the kinetic or the potential term, see {\em e.g.} \cite{gftsimp}. 

Geometrically, the role of simplicity constraints is to select a local $\SU(2)$ subgroup of $G$ at each point in a spatial hypersurface; this can be understood as `spontaneous symmetry breaking' by a field of {\em observers} with respect to which a local $\SU(2)$ subgroup is defined \cite{ssb}. A gauge-fixing with constant observer field leads to the Ashtekar--Barbero formulation in terms of $\SU(2)$; conversely, the Lorentz-covariant theory can be recovered from the $\SU(2)$ theory by specifying how it transforms under changes of observer.

In the GFT quantum analogue of this classical formalism, instead of starting with a larger gauge group $G$ and restricting representations, one can choose $G=\SU(2)$ and implement simplicity constraints through a choice of embedding map
\ben
\varpi: \; L^{2}(\SU(2)^4/\SU(2))  \rightarrow  L^{2}(H^4/H)
\een
where $H={\rm Spin}(4)$ or $H={\rm SL}(2,\bC)$. This embedding map is the analogue of a classical choice of observer at each point in a spatial hypersurface which determines an embedding of $\SU(2)$ into $G$. Its choice is not unique and different choices for $\varpi$ correspond to different spin foam models, see \cite{JHEP}. Here we assume that a suitable map $\varpi$ can be chosen, and one can work in Ashtekar--Barbero-type variables with $G=\SU(2)$.

Having established that an open connected subset of $\mathfrak{su}(2)^{\oplus 3}/{\rm Ad}_{\SU(2)}$ represents the space of nondegenerate homogeneous 3-metrics, we fix the coordinates 
\ben
B_{ij}:=-\half\,{\rm Tr}\left(B_i B_j\right)=\vec{B}_i\cdot\vec{B}_j\,,
\een
which is a global coordinate system. In the last equation, we have identified $\mathfrak{su}(2)\simeq\bR^3$ using the standard basis of Pauli matrices, {\em i.e.} $B=:\im\vec\sigma\cdot\vec{B}$ for $B\in\mathfrak{su}(2)$. 

In terms of the components of the spatial metric, using (\ref{def}), the function $B_{ij}$ corresponds to the minor of the entry $g_{ij}$ of the spatial metric $g$ defined by $g_{ij}=\delta_{AB}e_i^A e_j^B$:
\ben
B_{11}=g_{22}g_{33}-g_{23}^2\,,\quad{\rm etc.}
\een
We note that for $\det g\neq 0$, $g_{ij}$ is diagonal if and only if $B_{ij}$ is diagonal. $\det g$ can be computed by considering
\ben
\det B_{ij} = \frac{1}{6}\epsilon^{ijk}\epsilon^{lmn}B_{il}B_{jm}B_{kn} = (\det g)^2\,.
\een
Note that $g$, like $B_{ij}$, is by construction positive (semi-)definite, and $\det g\neq 0$ or $\det B_{ij}\neq 0$ is equivalent to the $B_i$ being linearly independent, and forming a basis of $\mathfrak{su}(2)\simeq\bR^3$. The space of non-degenerate matrices $B_{ij}$ is then again the homogeneous space ${\rm GL(3)}/ {\rm O}(3)\simeq {\rm SL(3)}/ {\rm O}(3)\times(\bR\backslash\{0\})$, where the $\bR\backslash\{0\}$ corresponds to an (oriented) overall volume factor, which we may restrict to be positive.

In these variables, having an isotropic universe, {\em i.e.} a 3-metric proportional to the identity matrix, is equivalent to $B_{ij}=a^2 \delta_{ij}$ for some $a^2>0$.

So far, we have treated $\mathfrak{su}(2)^{\oplus 3}/{\rm Ad}_{\SU(2)}$ only as a vector space, ignoring its Lie algebra structure. Using the basic commutation relations $\{\vec{B}_i,\vec{B}_j\}={\rm G}\delta_{ij}(\vec{B}_i\times\vec{B}_j)$,
\ben
\{B_{ij},B_{kl}\} = {\rm G}\,{\rm Tr}(B_1 B_2 B_3)(\epsilon_{ilj}\delta_{jk}+\epsilon_{jli}\delta_{ik}+\epsilon_{ikj}\delta_{jl}+\epsilon_{jki}\delta_{il})\,;
\een
the quotient $\mathfrak{su}(2)^{\oplus 3}/{\rm Ad}_{\SU(2)}$ inherits a non-commutative structure from $\mathfrak{su}(2)$. Expressed in terms of Lie algebra variables in $\mathfrak{su}(2)^{\oplus 3}/{\rm Ad}_{\SU(2)}$, effective quantum cosmology models derived from GFT condensates naturally describe a {\em non-commutative} quantum cosmology, with some similarity to models such as \cite{ncqc}. Here ${\rm G}$ is a dimensionful parameter which in LQG is normally a product of Newton's constant with a numerical factor, perhaps involving the Barbero--Immirzi parameter $\gamma$, and so corresponds to an inverse tension \cite{tension}. We also note that the coordinates $B_{ii}$ commute with all others; noncommutativity becomes only relevant for terms at least quadratic in anisotropies, and could be ignored if one linearises around isotropy.

To extend the discussion to connection variables, we need to choose a convenient set of coordinates on $\SU(2)^3/{\rm Ad}_{\SU(2)}$ which can be interpreted in terms of quantum cosmology. Recall that one interprets the elements of $\SU(2)$ as parallel transports of a gravitational connection which is taken as approximately constant over the scale of the tetrahedra,
\ben
g=:\mathcal{P}\exp\int_e\omega=\mathcal{P}\exp\int_0^\nu\;dx^i\,\omega_i\approx\exp(\nu\,\omega_x)\,,
\label{ident}
\een
if the coordinate system (on the spatial manifold) is chosen such that the edge $e$ has coordinate length $\nu$ in the $x$ direction. The adjoint action ${\rm Ad}_{\SU(2)}$ on $g$ then becomes $\omega_x\mapsto k\omega_x k^{-1}$, which corresponds to an $\SU(2)$ gauge transformation that is constant over $e$ (as is consistent with the assumption of constant $\omega$). Of course, as $\SU(2)$ is compact, and hence there are nonzero $T\in\mathfrak{su}(2)$ for which $\exp(T)=e$, there is no invertible map $g\mapsto\omega[g]$ that would allow a reconstruction of the `connection' $\omega$ from its parallel transports. At least in a neighbourhood of the identity in $\SU(2)$, we can write
\ben
g=\sqrt{1-\vec{\pi}[g]^2}\,{\bf 1}-\im\vec\sigma\cdot\vec{\pi}[g]\,,\quad|\vec\pi|\le 1\,,
\een
which defines coordinates $\vec{\pi}$ on $\SU(2)$. Comparing with (\ref{ident}), we have 
\ben
\vec\pi=-\vec\omega_x\sin(\nu|\vec\omega_x|)/|\vec\omega_x|\,,
\label{conn}
\een 
again using $\mathfrak{su}(2)\simeq\bR^3$, so that $\vec\pi$ corresponds to a `sine of the connection'. It is the natural variable arising from a discretisation of a (gravitational) connection, and replaces the connection in the holonomy corrections of loop quantum cosmology \cite{LivRev}. The adjoint action of $\SU(2)$ on itself acts as rotations of the `coordinate vector' $\vec\pi$. 

These coordinates are particularly useful if the Fourier transform (\ref{fourier}) is defined by $e_g(B):=\exp(\frac{\im}{2}\,{\rm Tr}(g B)/\hbar {\rm G})$, since this becomes simply $e_g(B)=\exp(\im\,\vec{\pi}[g]\cdot\vec{B}/\hbar {\rm G})$. This does not mean, of course, that the Fourier transform is the standard one on $\bR^3$: there is a non-trivial measure factor in (\ref{fourier}), ${\rm d}g={\rm d}\vec\pi(1-\vec\pi^2)^{-1/2}$. The coordinates $\vec\pi$ on the group and $\vec{B}$ on the Lie algebra are {\em not} canonically conjugate as phase space variables, and it is easy to see that such a coordinate choice on the group does not exist: for a phase space $T^*G=G\times\mathfrak{g}$, where $G$ is a non-Abelian Lie group, the Poisson bracket for the Lie algebra variables is induced by the Lie bracket,
\ben
\{B^{\alpha},B^{\beta}\}={c^{\alpha\beta}}_{\gamma}B^{\gamma}\,.
\een
A putative choice of coordinates $p^\alpha$ on the group such that $\{p^{\alpha},B^{\beta}\}\propto\delta^{\alpha\beta}$ is then not compatible with the Jacobi identity, as one would have
\ben
0\stackrel{?}{=}\{\{B^{\alpha},B^{\beta}\},p^{\gamma}\}+{\rm permutations}\propto c^{\alpha\beta\gamma}\,.
\een
From the same argument, one sees that the first correction to $\{p^{\alpha},B^{\beta}\}=c\,\delta^{\alpha\beta}+\ldots$ comes in at linear order in $p$. The coordinates $\vec\pi$ used in the following have this property: they are canonically conjugate to the $\vec{B}$ (including, for dimensional reasons, a factor of ${\rm G}$) up to terms of linear and higher order in $\vec\pi$ \footnote{For explicit forms of the Poisson brackets, see \cite{danielematti} where our coordinates $\vec\pi[g]$ are denoted $Y^i_e(g)$.}. In the definition of GFT condensates, we have assumed that gauge-invariant combinations of parallel transports are peaked on values close to the identity, which is a sufficient condition for guaranteeing that all components of the curvature remain small on the scale of the individual tetrahedra. This is the regime in which $|\vec\pi|\ll 1$ and $\vec\pi$ and $\vec{B}$ can be viewed as canonically conjugate.
 
We then choose the invariants under the left and right actions of $\SU(2)$ on $\SU(2)^4$
\ben
\pi_{ij}:= \vec\pi[g_i g_4^{-1}]\cdot\vec\pi[g_j g_4^{-1}]\,,
\een
with $|\pi_{ij}|\le 1$ and $\pi_{ii}\ge 0$, to define a coordinate system in a neighbourhood of the identity $[e,e,e]\in\SU(2)^3/{\rm Ad}_{\SU(2)}$. The coordinates $\vec{\pi}$ cover a hemisphere of $S^3\sim\SU(2)$ for each of the three copies of $\SU(2)$, mapping it to a three-ball $B^3\subset \bR ^3$, and $\pi_{ij}$ are invariant under the adjoint action of $\SU(2)$ acting as rotations of $\bR^3$. Note that the identity coset $[e,e,e]\in\SU(2)^3/{\rm Ad}_{\SU(2)}$ is in the boundary of $\SU(2)^3/{\rm Ad}_{\SU(2)}$, as it corresponds to a fixed point of ${\rm Ad}_{\SU(2)}$: $[e,e,e]=(e,e,e)$ \footnote{This is in the same way in which $r=0$ is in the boundary of $\bR^3/\SO(3)\simeq \bR^+$.}.

By (\ref{conn}), these coordinates correspond to gauge-invariant (in the sense explained above) combinations of the components of a `gravitational connection' as follows,
\ben
\pi_{ii}=\sin^2(\nu|\vec\omega_i|)\,,\quad \pi_{ij}=\cos\theta_{ij}\sin(\nu|\vec\omega_i|)\sin(\nu|\vec\omega_j|)\,,
\een
where $\theta_{ij}$ is the angle between the connection components $\vec\omega_i$ and $\vec\omega_j$ again viewed as elements of $\bR^3$. We will later be interested in the isotropic case where only $\pi_{ii}\neq 0$.

\section{The example: Laplace--Beltrami beyond the WKB approximation}

As an example of how effective quantum cosmology equations can be extracted from the dynamics of quantum gravity condensates, the discussion of \cite{PRL, JHEP} considered
\ben
\hat\mathcal{K}=\sum_{I=1}^4 \Delta_{g_I}+\mu\,,\quad \hat\mathcal{V}=0\,,
\label{lapl}
\een
where $\Delta_g$ is the Laplace--Beltrami operator on $\SU(2)\sim S^3$, and $\mu\in\bR$. The choice $\hat\mathcal{V}=0$ can arise from a condensate of the `dipole' type for which it was shown that, for a class of GFT potentials, the nonlinear term can be approximately neglected in the effective quantum cosmology equation. The choice of a Laplacian in the kinetic term can be motivated, among other arguments, by results in the renormalisation of GFT that suggest that such a kinetic term is generated by radiative corrections \cite{renorm}. Its presence is used to define a notion of scale for a renormalisation group flow in GFT \cite{renorm2}.

Let us consider (\ref{lapl}) as an example, and explicitly reduce the quantum cosmology equation from $\SU(2)^4$ to the variables $\pi_{ij}$ invariant under separate left and right actions of $\SU(2)$. It is immediate to see that (\ref{lapl}) satisfies (\ref{kinetic}), as the Laplace--Beltrami operator on $\SU(2)$ (defined with respect to the round metric on $S^3$) is bi-invariant. In terms of the coordinates $\vec\pi$ on $\SU(2)$, $\Delta$ is defined by
\ben
\Delta_g f(\pi[g]) = (\delta^{\alpha\beta}-\pi^{\alpha}\pi^{\beta})\partial_\alpha\partial_\beta f(\pi) - 3 \pi^{\alpha}\partial_\alpha f(\pi)\,.
\een
On functions on $\SU(2)^4/\SU(2)$ which can be identified as functions on $\SU(2)^3$, using the bijection $[g_1,g_2,g_3,g_4]\mapsto(g_1g_4^{-1},g_2g_4^{-1},g_3g_4^{-1})$, we can compute (in everything that follows, there is no summation convention for indices $i,j,k,l$ which run from 1 to 3)
\bena
&&\sum_{I=1}^4\Delta_{g_I}\sigma(g_1g_4^{-1},g_2g_4^{-1},g_3g_4^{-1})\nonumber
\\&=&\sum_i\Delta_{g_i}\sigma(g_1g_4^{-1},g_2g_4^{-1},g_3g_4^{-1})+\Delta_{g_4}\sigma(g_1g_4^{-1},g_2g_4^{-1},g_3g_4^{-1})\nonumber
\\&=&2\sum_i\left( (\delta^{\alpha\beta}-\pi_i^{\alpha}\pi_i^{\beta})\partial^i_\alpha\partial^i_\beta - 3 \pi_i^{\alpha}\partial^i_\alpha \right)\sigma(\vec\pi[g_i g_4^{-1}])\nonumber
\\&&+\sum_{i\neq j}\left(\left(\sqrt{1-\vec\pi_i^2}\sqrt{1-\vec\pi_j^2}+\vec\pi_i\cdot\vec\pi_j\right)\delta^{\alpha\beta}\right.\nonumber
\\&&\left.-{\epsilon^{\alpha\beta}}_{\gamma}\pi_i^{\gamma}\sqrt{1-\vec\pi_j^2}+{\epsilon^{\alpha\beta}}_{\gamma}\pi_j^{\gamma}\sqrt{1-\vec\pi_i^2}-\pi_j^{\alpha}\pi_i^{\beta} \right)\partial^i_\alpha\partial^j_\beta\sigma(\vec\pi[g_i g_4^{-1}])
\label{laplace}
\eena
where $\vec\pi_i:=\vec\pi[g_i g_4^{-1}]$. The Laplace--Beltrami operator with respect to $g_4$ gives two contributions, one which just doubles the other three contributions (which themselves directly follow from right-invariance of $\Delta_{g_i}$), and one which is `anisotropic'. (\ref{laplace}) is manifestly invariant with respect to the left $\SU(2)$ action $\vec\pi_i\mapsto O\vec\pi_i$.

We now take $\sigma$ to be a function on $\SU(2)^3/{\rm Ad}_{\SU(2)}$, {\em i.e.} a function that only depends on the coordinates $\pi_{ij}$, $\sigma(\vec\pi[g_i g_4^{-1}])=\sigma(\pi_{(ij)})$. As usual, round brackets denote symmetrisation, making explicit that there are really only six independent coordinates $\pi_{(ij)}$, as $\pi_{(12)}$ and $\pi_{(21)}$ refer to the same coordinate. Then, using the chain rule
\bena
\partial^i_\alpha\partial^j_\beta\sigma(\pi_{(ij)})&=&\sum_{k,l}\pi_{k\alpha}\pi_{l\beta}\frac{\partial^2\sigma}{\partial\pi_{(ik)}\partial\pi_{(jl)}}+\sum_{k}\pi_{i\alpha}\pi_{k\beta}\frac{\partial^2\sigma}{\partial\pi_{ii}\partial\pi_{(jk)}}\nonumber
\\&&+\sum_{k}\pi_{k\alpha}\pi_{j\beta}\frac{\partial^2\sigma}{\partial\pi_{(ik)}\partial\pi_{jj}}+\pi_{i\alpha}\pi_{j\beta}\frac{\partial^2\sigma}{\partial\pi_{ii}\partial\pi_{jj}}+\delta_{\alpha\beta}\frac{\partial\sigma}{\partial\pi_{(ij)}}\nonumber
\\&&+\delta_{\alpha\beta}\delta^{ij}\frac{\partial\sigma}{\partial\pi_{ii}}\,,
\eena
the action of $\sum_{I=1}^4\Delta_{g_I}$ on $\sigma$ becomes
\bena
&&\sum_{I=1}^4\Delta_{g_I}\sigma(\pi_{ij})\nonumber
\\&=&2\sum_{{i,k,l}\atop{i\not\in\{k,l\}}}(\pi_{(kl)}-\pi_{(ik)}\pi_{(il)})\frac{\partial^2\sigma}{\partial\pi_{(ik)}\partial\pi_{(il)}}+8\sum_{i\neq k}\pi_{(ik)}(1-\pi_{ii})\frac{\partial^2\sigma}{\partial\pi_{ii}\partial\pi_{(ik)}}\nonumber
\\&&+8\sum_i\pi_{ii}(1-\pi_{ii})\frac{\partial^2\sigma}{\partial\pi_{ii}^2}+4\sum_i(3-4\pi_{ii})\frac{\partial\sigma}{\partial\pi_{ii}}-4\sum_{i\neq j}\pi_{(ij)}\frac{\partial\sigma}{\partial\pi_{(ij)}}\nonumber
\\&&+\sum_{i\neq j}\left[\sum_{k,l}\left(\sqrt{1-\pi_{ii}}\sqrt{1-\pi_{jj}}\,\pi_{(kl)}+\pi_{(ij)}\pi_{(kl)}\right.\right.\nonumber
\\&&\left.-2\sqrt{1-\pi_{jj}}\pi_{[kli]}-\pi_{(jk)}\pi_{(il)}\right)\frac{\partial^2\sigma}{\partial\pi_{(ik)}\partial\pi_{(jl)}}
\label{lapexp}
\\&&+2\sum_k \left(\sqrt{1-\pi_{ii}}\sqrt{1-\pi_{jj}}\,\pi_{(ik)}-\pi_{[ijk]}\sqrt{1-\pi_{ii}}\right)\frac{\partial^2\sigma}{\partial\pi_{ii}\partial\pi_{(jk)}}\nonumber
\\&&+\left.\sqrt{1-\pi_{ii}}\sqrt{1-\pi_{jj}}\,\pi_{ij}\frac{\partial^2\sigma}{\partial\pi_{ii}\partial\pi_{jj}}+3\sqrt{1-\pi_{ii}}\sqrt{1-\pi_{jj}}\,\frac{\partial\sigma}{\partial\pi_{(ij)}}\right]\,.\nonumber
\eena

This is now expressed only in terms of the coordinates $\pi_{(ij)}$, except for $\pi_{[ijk]}:=\vec\pi_i\cdot(\vec\pi_j\times\vec\pi_k).$ Up to a sign, this can be reconstructed from
\ben
\pi_{[ijk]}^2=\det\begin{pmatrix}
{\pi_{ii} & \pi_{ij} & \pi_{ik} \cr \pi_{ij} & \pi_{jj} & \pi_{jk} \cr \pi_{ik} & \pi_{jk} & \pi_{kk} }
\end{pmatrix}\,.
\label{det}
\een
This sign is a choice of orientation of the three vectors $\{\vec\pi_i\}$ which can not be obtained from the ${\rm O}(3)$ invariant combinations $\vec\pi_i\cdot\vec\pi_j$. As above, the space of non-degenerate matrices $\{\pi_{ij}\}$, for which the determinant in (\ref{det}) is non-vanishing, is ${\rm GL(3)}/ {\rm O}(3)\simeq {\rm SL(3)}/ {\rm O}(3)\times(\bR\backslash\{0\})$ which splits into two connected components. The coordinates $\pi_{ij}$ only parametrise one of these, and we can choose the component with $\pi_{[123]}>0$.

(\ref{lapexp}) is the explicit expression of $\sum_{I=1}^4\Delta_{g_I}\sigma(\pi_{ij})$ in the coordinates $\pi_{(ij)}$ on $\SU(2)\backslash\SU(2)^4/\SU(2)$. Substituting (\ref{lapexp}) into $(\sum_{I=1}^4\Delta_{g_I}+\mu)\sigma(\pi_{(ij)})=0$ defines a homogeneous, anisotropic quantum cosmology model for an empty universe without matter. As suitable explicit solutions to it are difficult to construct, the strategy adopted in \cite{JHEP} was to go to a WKB limit: assume that
\ben
\sigma(\pi_{(ij)}) = A(\pi_{(ij)})\exp\left(\im S(\pi_{(ij)})/\hbar {\rm G}\right)
\een
where $S(\pi_{(ij)})$ oscillates rapidly compared to $A(\pi_{(ij)})$, take the limit of $\hbar {\rm G}\rightarrow 0$, define the momentum $B^{(ij)}=\partial S/\partial\pi_{(ij)}$ conjugate to $\pi_{(ij)}$, and interpret the resulting classical equation for $\pi_{(ij)}$ and $B^{(ij)}$ in terms of gravitational dynamics. In the isotropic case, this led to a holonomy-corrected Friedmann equation for vacuum Riemannian gravity,
\ben
P^2 - k = O(\hbar{\rm G})\,,
\label{fried}
\een
where $P$ was identified with $\sin(\nu\,\omega)$ for the gravitational connection $\omega$, for some $\nu\in\bR$ \footnote{If we (tentatively) identify $\omega$ with the Ashtekar--Barbero conncetion, in the LQC notation of \cite{LivRev} we would have $P=\sin(\delta\,c)$ where $\delta$ can be a constant or depend on phase space variables. We will investigate the precise relation of the variables $(\pi_{(ij)},B^{(ij)})$ to the phase space variables of LQC more closely in future work, making such identifications precise.}, and $k>0$ is a constant interpreted as spatial curvature. This effective Friedmann equation seemed to have no solutions, as a consistent approximation of a continuum metric by discrete building blocks seemed to require $P\ll 1$, while $k$ is a dimensionless ratio of WKB variables depending on the state and of $O(1)$. 

If (\ref{fried}) were the Friedmann equation for continuum variables, one could change coordinates to rescale $k$, but this freedom is not present here: the geometric variables are expressed with respect to the scale given by the discrete building blocks of geometry. While in the identification $P=\sin(\nu\,\omega)$ both $\nu$ and $\omega$ depend on a choice of coordinates, $P$ does not; its value is fixed by properties of the tetrahedra in the condensate. The issue that $P\ll 1$ contradicts (\ref{fried}) cannot be resolved by changing coordinates.

Here we address two possible issues with the WKB analysis. The first is that the assumption of semiclassicality excludes many potentially interesting solutions. Generic quantum gravity condensates are not semiclassical at all, and it may not be meaningful to look for states that have semiclassical properties at the Planck scale. Large-scale observables should display semiclassical behaviour to agree with what we see, but this requirement may be different from the very simple WKB criterion on the condensate wavefunction, which would impose semiclassical behaviour already on the microscopic degrees of freedom. Our results in this section support the viewpoint, coming also from full quantum gravity, that the latter is not a physically meaningful assumption.

The second issue is technical. The WKB approximation in \cite{JHEP} was done at the level of $\SU(2)^4$, with the symmetries of the wavefunction translated into relations among the WKB variables, which were then substituted into the WKB equations. It is not clear whether this is equivalent to first reducing the quantum equation by using symmetries and then only introducing WKB variables for the gauge-invariant quantities.

To extend the analysis of \cite{PRL, JHEP}, we avoid a WKB approximation, and derive analytical solutions to the quantum equation $(\sum_{I=1}^4\Delta_{g_I}+\mu)\sigma(\pi_{(ij)})=0$. We consider isotropic states for which $\sigma$ only depends on the coordinates $\pi_{ii}$. Then from (\ref{lapexp}),
\bena
\sum_{I=1}^4\Delta_{g_I}\sigma(\pi_{(ij)})&=&8\sum_{i}\pi_{ii}(1-\pi_{ii})\frac{\partial^2\sigma}{\partial\pi_{ii}^2}+4\sum_i(3-4\pi_{ii})\frac{\partial\sigma}{\partial\pi_{ii}}\nonumber
\\&&+4\sum_{i\neq j}\sqrt{1-\pi_{ii}}\sqrt{1-\pi_{jj}}\,\pi_{(ij)}\frac{\partial^2\sigma}{\partial\pi_{jj}\partial\pi_{ii}}\,.
\eena
Due to the appearance of off-diagonal $\pi_{(ij)}$ in the second term, the only solution to $(\sum_{I=1}^4 \Delta_{g_I}+\mu)\sigma(\pi_{ii})=0$ compatible with our ansatz seems to be
\ben
\sigma(\pi_{ii})=\sigma_1(\pi_{11})+\sigma_2(\pi_{22})+\sigma_3(\pi_{33})
\label{salj}
\een
with all $\sigma_i$ separately satisfying
\ben
2p(1-p)\sigma_i''(p)+(3-4p)\sigma_i'(p)+\mu_i\sigma_i(p)=0
\een
for some $\mu_i$ such that $\sum_i\mu_i=\frac{1}{4}\mu$. Hence, all solutions that only depend on diagonal elements $\pi_{ii}$ can be obtained as a sum of solutions to the ordinary differential equation
\ben
2p(1-p)\sigma''(p)+(3-4p)\sigma'(p)+\hat\mu\,\sigma(p)=0\,.
\label{ode}
\een
Since (\ref{salj}) completely decouples $\pi_{11}$, $\pi_{22}$ and $\pi_{33}$, let us now set $\sigma_2=\sigma_3=0$, and leave $\hat\mu$ arbitrary (one could set $\hat\mu=\frac{1}{4}\mu$). We can then first compare (\ref{ode}) to the Friedmann equation obtained in the WKB limit in \cite{PRL, JHEP}. The WKB limit of (\ref{ode}) gives simply
\ben
2p(1-p) = O(\hbar {\rm G})\,,
\een
and hence $p\approx 0$ or $p\approx 1$. $p$ can be identified with $\sin^2(\nu\,\omega)$, and so again this is a vacuum Friedmann equation, either for a flat or for a closed universe. Only $p\approx 1$ is compatible with the previous result (\ref{fried}); the solution $p\approx 0$, describing a flat universe, appears when taking the WKB limit only for isotropic, gauge-invariant variables.

We can now compare this approximation to the explicit general solution to (\ref{ode}), 
\ben
\sigma(p)=\chi \sqrt[4]{\frac{1-p}{p}}\,P_{\half(\sqrt{1+2\hat\mu}-1)}^{\half}(2p-1) + \upsilon \sqrt[4]{\frac{1-p}{p}}\,Q_{\half(\sqrt{1+2\hat\mu}-1)}^{\half}(2p-1)\,,
\label{gensol}
\een
where $P_n^m(x)$ and $Q_n^m(x)$ denote associated Legendre functions of the first and second kind, respectively, and $\chi$ and $\upsilon$ are constants.

One might worry about regularity at $p=0$ and $p=1$. Asymptotically as $p\rightarrow 0$,
\ben
\sqrt[4]{\frac{1-p}{p}}\,P_{\half(\sqrt{1+2\hat\mu}-1)}^{\half}(2p-1)\sim\frac{\cos\left(\frac{\pi}{2}\sqrt{1+2\hat\mu}\right)}{\sqrt{\pi}\,\sqrt{p}}
\een
whereas the solution is finite at $p=1$. If the cosine has a zero, {\em i.e.} $\hat\mu=2N(N+1)$ for non-negative integer $N$, the function approaches a constant as $p\rightarrow 0$.

For the second branch, near $p=0$
\ben
\sqrt[4]{\frac{1-p}{p}}\,Q_{\half(\sqrt{1+2\hat\mu}-1)}^{\half}(2p-1)\sim -\frac{\sqrt{\pi}\sin\left(\frac{\pi}{2}\sqrt{1+2\hat\mu}\right)}{2\sqrt{p}}\,,
\een
so that the function is finite as $p\rightarrow 0$ only if $\hat\mu=2N^2-\half=2(N+\half)(N-\half)$ for integer $N$. The solution goes to zero at $p=1$ as $\sqrt{1-p}$. 

If one were looking for solutions that are regular on the 3-sphere, only one branch and only in the cases $\hat\mu=2N(N+1)$, for half-integer $N$, would be admissible. These are the spherically symmetric eigenmodes of the Laplacian on $S^3$ that are usually considered (see {\em e.g.} \cite{laplref}). However, here we only require $\sigma(p)$ to be normalisable with respect to the Hilbert space measure for effective quantum cosmology wavefunctions. This measure is the one induced from the full quantum gravity Fock space, and depends on how precisely the condensate is defined, as $\sigma(p)$ should correspond to a normalisable condensate state in the GFT Fock space. In the simplest case that we will assume here, $\sigma$ defines a `single-particle' condensate of the form (\ref{singcon}). Then the criterion for normalisability is 
\ben
\int ({\rm d} g)^3 \;|\sigma(g_1,g_2,g_3)|^2<\infty\,,
\label{normalis}
\een
and the Hilbert space of condensate wavefunctions $\sigma$ is $L^2(\SU(2)^3/{\rm Ad}_{\SU(2)})$. Note that wavefunctions need not be normalised to one, as the integral (\ref{normalis}) gives the average total particle number in the Fock space. For functions $\sigma$ that just depend one coordinate $p=\pi_{ii}$, the normalisability condition reduces to
\ben
\int \frac{{\rm d}\vec\pi}{\sqrt{1-\vec\pi^2}}\;|\sigma(\pi_{ij})|^2 = 2\pi \int\limits_0^1 dp\;\frac{\sqrt{p}}{\sqrt{1-p}}|\sigma(p)|^2<\infty\,.
\label{norma}
\een
With respect to this measure, the general solutions (\ref{gensol}) are always normalisable, for any value of $\hat\mu$. They are analytical solutions to the effective quantum cosmology model that correspond to homogeneous, isotropic universes, which in general do not display any form of semiclassical behaviour. Generic solutions, in particular all solutions if $\hat\mu$ is not of the form $\hat\mu=2N(N+1)$, diverge as $p\rightarrow 0$, as does the probability density \footnote{This notion of `probability density' is defined simply with respect to a measure on the Hilbert space; we do not suggest an operational measurement interpretation for the wavefunction of an empty universe.} $\frac{\sqrt{p}}{\sqrt{1-p}}|\sigma(p)|^2$. For the first branch of solutions, there can also be a divergence in $\frac{\sqrt{p}}{\sqrt{1-p}}|\sigma(p)|^2$ as $p\rightarrow 1$, but for the second branch the probability density remains finite. 

The detailed shape of these solutions determines whether they describe condensates of tetrahedra satisfying with high probability $p\ll 1$, {\em i.e.} the assumption of very small curvature, relative to the scale of the tetrahedra, that was made in the analysis of \cite{JHEP}.

\begin{figure}
\begin{center}
\includegraphics[scale=0.6]{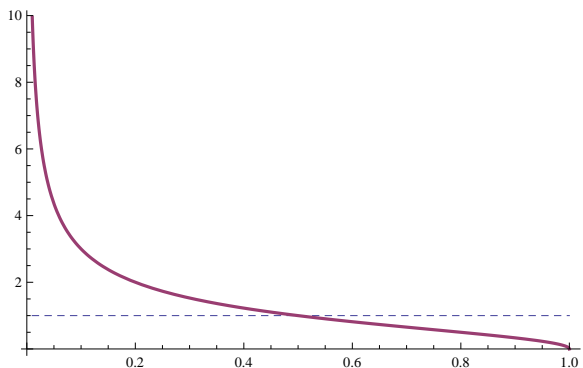}
\caption{Probability density defined by the two independent solutions for $\hat\mu=0$.}
\label{fig1}
\end{center}
\end{figure}

Let us look at some specific choices for $\hat\mu$. First, we take $\hat\mu$ to be non-negative. If $\hat\mu=2N(N+1)$ for some non-negative integer $N$, the first branch of solutions simply becomes a polynomial in $p$. In the simplest case where $\hat\mu=0$, this is just a constant (Fig. \ref{fig1}; in all of these plots the first branch is plotted in dashed blue while the second branch is thick red). The second branch is clearly peaked near $p=0$.

\begin{figure}
\begin{center}
\includegraphics[scale=0.6]{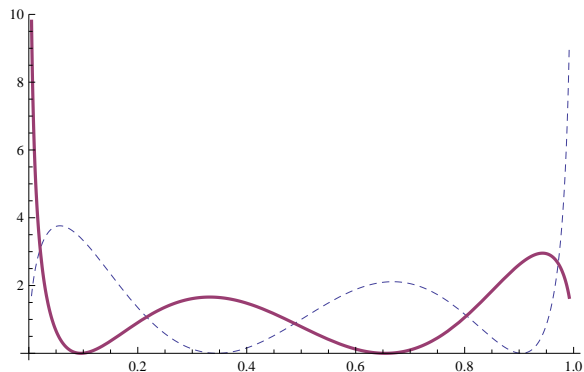}
\caption{Probability density defined by the two independent solutions, $\hat\mu=12$.}
\label{fig2}
\end{center}
\end{figure}

For $\hat\mu=12$, the respective probability densities are shown in Fig. \ref{fig2}. While they are maximal near $p=1$ and $p=0$, respectively, the distributions they define are broad, and not clearly compatible with assuming that most tetrahedra should have $p\ll 1$.

\begin{figure}
\begin{center}
\includegraphics[scale=0.6]{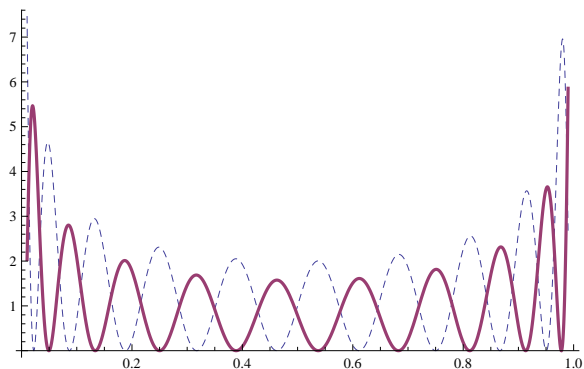}
\caption{Probability density defined by the two independent solutions, $\hat\mu=220$.}
\label{fig7}
\end{center}
\end{figure}

Oscillating solutions arise only for large enough positive values for $\hat\mu$. We show the two probability densities for $\hat\mu=220$ in Fig. \ref{fig7}. Both define rather broad probability distributions, incompatible with assuming $p\ll 1$. This is due to the measure in (\ref{norma}), as $|\sigma(p)|^2$ alone is peaked near $p=0$ in both cases; conclusions drawn from the WKB limit can be modified when one uses the proper type of inner product.

\begin{figure}
\begin{center}
\includegraphics[scale=0.6]{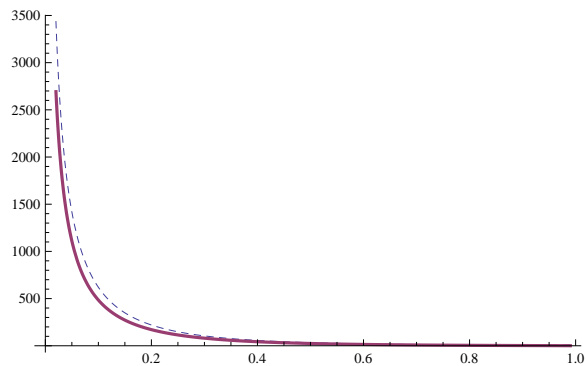}
\caption{Probability density defined by the two independent solutions, $\hat\mu=-4$.}
\label{fig3}
\end{center}
\end{figure}

\begin{figure}
\begin{center}
\includegraphics[scale=0.6]{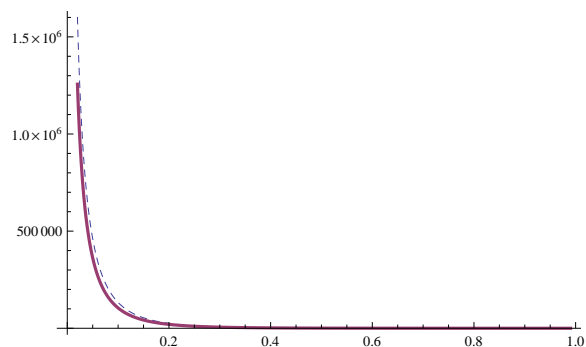}
\caption{Probability density defined by the two independent solutions, $\hat\mu=-12$.}
\label{fig4}
\end{center}
\end{figure}

For negative values of $\hat\mu$, both branches of solutions are strongly peaked on values close to zero. We give the probability densities for $\hat\mu=-4$ in Fig. \ref{fig3} and for $\hat\mu=-12$ in Fig. \ref{fig4}. For $\hat\mu<0$, the model can be claimed to be predict a condensate of almost flat tetrahedra. Ultimately, the fundamental GFT model determines whether positive or negative $\hat\mu$, or $\hat\mu\gg1$, should be considered (in many examples such as \cite{renorm}, $\hat\mu<0$). 

By redoing the WKB analysis of \cite{PRL, JHEP} and by analytical computation of simple solutions of the effective quantum cosmological dynamics, we have shown that the WKB argument which was seemingly in contradiction with having near-flat tetrahedra cannot be trusted. The solutions we found either do not oscillate rapidly or do not agree with the WKB result, due to the choice of inner product. The solutions for $\hat\mu>0$ define broad distributions while for $\hat\mu<0$ they are peaked near $p=0$. They are normalisable for any value of $\hat\mu$ if one takes the wavefunction $\sigma$ as defining a single-particle condensate (\ref{singcon}), even though they are not regular eigenmodes of the Laplacian on the 3-sphere. 

\section{Discussion}

Condensate states in group field theory can be used to derive effective quantum cosmology models directly from a proposal for the dynamics of a quantum theory of discrete geometries. We have illustrated the interpretation of the configuration space of gauge-invariant geometric data of a tetrahedron, the domain of the condensate wavefunction, as a minisuperspace of spatially homogeneous 3-metrics. 

The approach taken here is very different from the more conventional one of quantising only classical degrees of freedom that remain after imposing a symmetry. It makes assumptions about the approximate form of a fully dynamical quantum gravity state, similar to the assumptions one makes when treating interacting quantum systems in condensed matter physics. The validity of these assumptions can be verified; for instance, one can compute fluctuations around the mean field given by the condensate wavefunction $\sigma(g_I)$ and see whether they remain small. From the classical interpretation of the geometric data associated to these states, one expects that they are good approximations as long as the curvature remains small on the scale of the tetrahedra \cite{JHEP}. Again, this assumption can be verified by analysing the effective quantum cosmology equations, and in \cite{PRL, JHEP} there seemed to be a tension as the WKB approximation indicated that the curvature was peaked at large (presumably Planckian) values. This was our main motivation for revisiting the model of \cite{PRL, JHEP}. We found that a more consistently derived WKB approximation has a second solution corresponding to flat universes. Then, rather than assuming semiclassical properties for the condensate wavefunction, we gave simple isotropic solutions to the full quantum cosmology equation. These solutions depend on the `mass parameter' $\mu$. For negative $\mu$, they violate the assumptions of the WKB approximation, but are peaked on small curvature $p\ll 1$ and thus consistent with the expectations of the classical picture, as well as with the Friedmann equation $p\approx 0$. For positive $\mu$ states show a wider distribution of curvature.

The effective Friedmann equation (\ref{fried}), as discussed in \cite{PRL, JHEP}, came out of the WKB approximation of (\ref{effeq}) with (\ref{lapl}). Once it is accepted that a WKB-type condensate wavefunction may not give a physically relevant approximation to the dynamics, one might ask whether (\ref{lapl}) still corresponds to an interesting model of quantum cosmology. The explicit examples given in the paper show that this depends strongly on the value of the `mass parameter' $\mu$ in the fundamental theory; this parameter dropped out in the WKB limit. For negative $\mu$ solutions are strongly peaked near $p=0$, and (\ref{lapl}) implements the Friedmann equation $p=0$ describing a pure vacuum, spatially flat universe. In any case, (\ref{lapl}) remains a useful example to consider, because it is simple enough for explicit solutions to be constructed, so that the physical interpretation of GFT condensates and their cosmology can be discussed. Further work will be required to conclusively answer whether the model can reproduce some features of general relativity.

In the reduction to isotropic states, the model we have studied is fully constrained: there is only one degree of freedom (essentially given by the scale factor) and one constraint. One could add anisotropies or include matter degrees of freedom into the model. For instance, a massless scalar field can be introduced \cite{JHEP} by taking
\ben
\hat\mathcal{K}=\sum_{I=1}^4 \Delta_{g_I}+\tau\,\frac{\partial^2}{\partial\phi^2}+\mu
\label{knew}
\een
for an extended GFT model with a field on $G^4\times\bR$ where $\bR$ parametrises the scalar field. Adopting this prescription and decomposing $\sigma(p,\phi)=\sum_{\omega}\sigma_\omega(p)e^{\im\omega\phi}$, general isotropic solutions would be superpositions of the solutions given above with $\hat\mu=\frac{1}{4}(\mu-\tau\omega^2)$, and one could try to construct wavepackets similar to \cite{kiefer}. Requiring these to be composed out of rapidly oscillating modes would require choosing $\tau<0$ and/or a restriction on the values of $\omega$, depending on the value of $\mu$. The physical meaning of these conditions and of the choice (\ref{knew}) from the viewpoint of quantum gravity is however rather unclear.

We conclude that the criterion of {\em semiclassicality} for condensate states describing quantum cosmology has to be phrased more carefully to justify results such as an effective Friedmann equation (\ref{fried}) that can support the potential usefulness of the choice (\ref{lapl}) for quantum cosmology. It is only for large-scale observables, such as the total volume (of the universe), that semiclassical behaviour is required. The condensate wavefunction itself captures the properties of what presumably describes a highly quantum-mechanical many-particle state of Planck-scale objects. It carries much more information than a usual quantum cosmology wavefunction, {\em e.g.} about correlations between different quanta or about the scaling of geometric observables with the particle number. Using this information will be necessary for adding inhomogeneities \cite{JHEP}, and for potentially making contact with CMB observations. All of this motivates further systematic studies of the quantum cosmology of (loop) quantum gravity condensates.

\section*{Acknowledgements}
I thank Daniele Oriti and the referees for helpful and valuable comments on the manuscript. Research at Perimeter Institute is supported by the Government of Canada through Industry Canada and by the Province of Ontario through the Ministry of Research \& Innovation.

%%%%%%%%%%%%%%%%%%%%%%%%%%%%%%%%%%%%%%%%%%%%%%%%%%%%%%%%%%%%%%%%%%%%%%%%%%%%%%%%%%%%%%%%%%%%%%%%
%%%%%%%%%%%%%%%%%%%%%%%%%%%%%%%%%%%%%%%%%%%%%%%%%%%%%%%%%%%%%%%%%%%%%%%%%%%%%%%%%%%%%%%%%%%%%%%%

\end{document}